\documentclass[a4paper,11pt, useAMS]{article}
\pdfoutput=1 

\usepackage{jcappub} 

\usepackage[T1]{fontenc} 

\title{\boldmath First constraints on Fuzzy Dark Matter from the dynamics of stellar streams in the Milky Way}

\author[a,b]{Nicola C. Amorisco}
\author[a]{, A. Loeb}

\affiliation[a]{Institute for Theory and Computation,  Harvard-Smithsonian Center for Astrophysics,  60 Garden St.,  MS-51,  Cambridge,  MA 02138,  USA}
\affiliation[b]{Max Planck Institute for Astrophysics,  Karl-Schwarzschild-Strasse 1, 85748 Garching, Germany}

\emailAdd{nicola.amorisco@cfa.harvard.edu}

\abstract{We present a novel method to constrain the mass of ultra-light bosons as the dark matter 
using stellar streams formed by disrupting Globular Clusters in the Milky Way. The turbulent density field
of Fuzzy Dark Matter (FDM) haloes results in perturbations and dynamical heating of thin streams.
Using numerical simulations based on an effective model, we explore the magnitude of this phenomenon 
and show that this is observable for the range of axion masses $m_a$ that is interesting for alleviating the 
`small-scale problems' of $\Lambda$CDM. We derive an analytical model for the thickening 
of thin stellar streams and obtain an early conservative lower limit for the boson mass of $m_a > 1.5 \times 10^{-22}~$eV, 
using pre-Gaia literature data for six Milky Way streams and after marginalizing over physical parameters.
This demonstrates the great promise for using this novel dynamical method as a fully independent
probe of FDM, to complement results based on Lyman-$\alpha$ forest data.}

\begin{document}
\maketitle
\flushbottom

\section{Introduction}
\label{sec:intro}

Ultra-light bosons \citep{Peccei,Wilczek,WeinbergE} have been proposed as a viable alternative to cold dark matter (CDM) \cite{Spergel,HuBarkana,Matos,Peebles,Goodman}.
A bosonic scalar field is compatible with current cosmological constraints \cite{Li,Hlozek,Hlozek2}, as it behaves like CDM on large scales.
However, the two models differ at non-linear scales for axion masses $m_a\lesssim 10^{-21}~$eV, 
corresponding to a de Broglie wavelength of $\sim1~$kpc.
Differences include a suppression in the small-scale power spectrum \cite{Frieman,Amendola,Marsh} 
and the formation of centrally cored dark matter haloes \cite{Lee,Guzman}, 
both of which could ease some of the small-scale challenges of galaxy formation in CDM \cite{MarshR,Hui}. 
For instance, an axion mass of $m_a\lesssim 2\times10^{-22}~$eV would reproduce the observed kinematics 
of multiple stellar sub-populations in the Milky Way dwarf Spheroidal satellites \cite{Pop,Chen,GonzalezMorales},
without the need to invoke the strong stellar feedback required in CDM \cite{Weinberg,Bullock}. 
As such, Fuzzy Dark Matter (FDM) has recently attracted attention as a viable alternative to CDM. 

A number of analytic studies have addressed the properties of dark matter haloes in FDM, predicting the formation 
of a stable self-gravitating Bose-Einstein condensate characterized by a central constant-density core \cite{Gleiser,Sin,Chavanis}.
Recent numerical studies \cite{Schive,moczTB} have confirmed this prediction and showed that, as in CDM, the density profile of FDM haloes is 
well fit by a Navarro-Frenk-White profile \cite[NFW,][]{NFW}, in which the most central region of the cusp is replaced by the soliton core. 
The size of the latter varies with both halo mass $M_{vir}$ and $m_a$ \cite{Schive,moczTB}: for a Milky Way sized halo with $M_{vir}=10^{12} M_\odot$, 
it is $r_{sc}\sim 0.2/m_{22}~$kpc, where $m_a=m_{22}\times10^{-22}~$eV. 

Both analytic work and numerical simulations have shown that FDM haloes are characterized by a sustained turbulent behavior, 
with order unity oscillations in the density field \cite{moczSP,Schive,moczTB}. These fluctuations, in the form of soliton-sized 
clumps \cite{Schive}, are caused by the reconnection of quantum vortex lines \cite{moczTB}, a phenomenon that is shared by 
Bose-Einstein condensates in absence of self-gravity \cite{Kobayashi,Baggaley}.
Turbulence may cause observable dynamical heating, offering means to constrain $m_a$. However, Ref. \cite{Hui} finds that 
the timescales for {\it(i)} the disruption of binary stars, {\it(ii)} the thickening of the Galactic disk and  {\it(iii)} the heating of open 
and globular clusters (GCs), are exceedingly long for these processes to provide useful constraints\footnote{Except possibly 
for the case of GCs close to the Galactic center.}. Here, we explore the dynamical heating of stellar streams 
of disrupting GCs, using both analytic and numerical methods. We set out to investigate what is the magnitude of the 
effect of quantum turbulence on the dynamics of GC streams in the interesting range of axion masses. 
We find that thin, kinematically cold stellar streams in the Milky Way represent a promising 
dynamical probe of the axion mass, providing means to set local dynamical constraints to this alternative dark matter model.

In Section 2 we derive an analytical model for the dynamical heating of stellar streams, which we test using 
effective numerical simulations in Section 3. In Section 4 we use pre-Gaia literature data for a selection of Milky Way GC
to illustrate the promise of this novel method and to derive a first constraint on the axion mass.

\section{Dynamical heating of cold stellar streams}

Due to their internal dynamical coherence (internal velocity dispersion $\lesssim1$~km~s$^{-1}$ \cite{OdenkirchenK,Koposov}) 
GC streams are excellent gravitational detectors of dark matter subhaloes \cite{Ibata,Johnston,Carlberg,ErkalP5, Bovy1, Bertone} and other baryonic 
disturbances \cite{AmoriscoGMC,Hattori,Pearson}. In FDM, GC streams experience the additional perturbing effects of quantum fluctuations. 

There are two qualitative differences between the disturbances due to the flyby of dark matter subhaloes in a $\Lambda$CDM halo 
and those due to FDM granules. 
\begin{itemize}
\item{First, the mass fraction in bound subhaloes in the 
central regions of a Milky Way sized halo is of only a few per cent \cite{Springel, GarrisonKimmel,Sawala,Diemand},
implying a limited number of encounters per stream \cite{ErkalNS}. It follows that the strongest observable perturbations 
-- in the form of density `gaps' and associated kinematic disturbances -- are due to the 
rare encounters with the most massive subhaloes, a process that is stochastic in nature. Instead, when orbiting a FDM halo, 
streams experience the continuous, recurring perturbations of quantum turbulence. For axion masses 
$m_{22}\gtrsim1$, each single encounter with a soliton-sized
quantum density fluctuation has a limited perturbing effect. However, since all of the dark 
matter participates in the turbulence (rather than just a small fraction), the rate of flybys is high.
This results in an inevitable process of diffusive heating.}
\item{The second qualitative difference 
has to do with the opposite dependences of the strength of the perturbing effects with 
galactocentric radius.While $\Lambda$CDM subhaloes become rarer in the innermost regions of the 
halo \cite{Springel, GarrisonKimmel,Sawala} making stochasticity ever more important, in FDM perturbations increase with decreasing radius, 
as the order-unity density fluctuations scale with the local dark matter density itself \cite{Hui}.}
\end{itemize}

Simulating a Milky-Way sized FDM halo self-consistently and resolving the dynamics 
of individual quantum clumps is extremely challenging at the moment, especially for the 
range of axion masses that is most interesting, $m_{22}\gtrsim 0.1$. Even more difficult would 
be to also resolve the internal dynamics of a GC and its stream in such a self-consistent FDM simulation. 
Therefore, here we provide a first quantification of the effect of quantum turbulence based on an effective 
description of the turbulent density field of FDM haloes. 

\subsection{An effective model: `quasiparticles'}

In order to quantify the dynamical heating of a thin stellar stream, we follow Ref. \cite{Hui} and adopt 
a phenomenological model of the density fluctuations in FDM haloes, describing quantum clumps with `quasiparticles'. 
We base this model on the results of recent self-consistent simulations \cite{Schive,moczTB}: quasiparticles 
have a mass 
\begin{equation}
M_{eff}=\rho V_{cl}=4\pi\rho r_{sc}^3/3\ , 
\label{meff}
\end{equation}
with $\rho$ being the time-averaged local dark matter density and $r_{sc}$ being the size of the halo's soliton core, $r_{sc}= r_{sc,1}/m_{22}~$kpc.
Using the soliton radius appropriate for a Milky Way sized halo, $r_{sc,1}=0.2~$kpc, 
Eqn.~(\ref{meff}) implies that the mass of quantum clumps at the solar radius  ($\rho_\odot=10^7 M_\odot$~kpc$^{-3}$, \cite{Read}) is
\begin{equation}
M_{eff,\odot}= 3.4\times10^5\ m_{22}^{-3}\ M_\odot\ ,
\label{meffr}
\end{equation}
showing the steep dependence on the axion mass. We assume that effective FDM clumps have a 
smooth density profile, represented by Plummer spheres with a smoothing length $\epsilon=r_{sc}$, 
and that they move within the halo potential with velocities appropriate for an isotropic 
system in dynamical equilibrium.

The choices above should be intended as temporary working assumptions, broadly guided by the results of currently available self-consistent 
FDM simulations. For instance, we are using that all quantum clumps at some given radius have the same effective mass $M_{eff}$, 
while it is likely that a distribution of masses would be more appropriate. In turn, such a distribution might also include
secondary dependences on galactocentric radius, or on clump velocity. 
Future self-consistent simulations will provide a better characterisation of the sustained turbulent density field of FDM halos, 
allowing for a more faithful description of the spectrum of effective masses, clump velocities, smoothing radii, as well as of any correlations
between these properties. It is straightforward to take these into account within the framework of the analytic model we are about to set out.

\subsection{An analytical model of the stream thickness}

We consider a stellar stream with orbit $r(t)$ and we wish to describe the thickening caused 
by the continuous encounters with the FDM clumps. We consider the conservative case 
of a spherically symmetric underlying gravitational potential, since this minimizes the growth 
rate of the stream thickness \cite{Helmi,ErkalSSS}. The dynamics of tidal streams within their 
orbital plane and perpendicularly to it are qualitatively different, therefore we treat these two 
degrees of freedom separately. 
We use the symbol $\omega_\parallel $ to indicate the angular width of the stream 
in the orbital plane and $\omega_z$ for the angular width perpendicularly to the orbital plane.
The observed angular width of any given stream on the sky can be obtained as
a combination of the two, which of course depends on the viewing angle. 

\subsubsection{Vertical angular width}
The thickness of a tidal stream in the direction perpendicular to the orbital plane, $\mathbf{\hat z}$, 
is due to the misalignment of the orbital planes of its member stars \cite{JohnstonS,AmoriscoS,ErkalSSS}. 
Encounters with the quantum granules cause additional scattering of such individual planes.
Following an encounter and a velocity change $\mathbf{\delta v}$, a star's orbital plane is tilted by the angle 
\begin{equation}
\delta\theta=\arctan \left( \delta v_z/v_\phi\right) \approx\delta v_z/v_\phi\ ,
\label{ztilt}
\end{equation}
where $v_\phi$ is the angular velocity of the stream at the location of the encounter.   
We assume that the process is a diffusive one, i.e. that the number of encounters is large enough 
that the mean orbital plane of the stream is unchanged and that individual encounters
are uncorrelated. In this case FDM turbulence causes a vertical thickening rate of
\begin{equation}
{{d\langle\omega^2_z\rangle}\over{dt}} \approx \left\langle {\iint} f(r,\mathbf{v_{cl}})d^3\mathbf{v_{cl}}   \ 2\pi b db\ n\ v_{rel} \left({{\delta v_z}\over{v_\phi}}\right)^2\right\rangle_{r(t)}\ ,
\end{equation}
where $f(r,\mathbf{v_{cl}})$ is the distribution of clump velocities at radius $r$, $b$ is the impact parameter, $n=1/V_{cl}$
is the number density of clumps, $v_{rel}$ is the relative velocity between star and clump. 
All of $v_{rel}$, $\delta v_z$ and $v_\phi$ depend on the proper velocity of the clump $\mathbf{v_{cl}}$ and on the specific orbit of 
the stream, $r(t)$, at the location of the encounter. We use the symbol $\langle\cdot\rangle_{r(t)}$ to indicate the average 
over one radial orbital period.
Making use of the impulse approximation and absorbing all dependences on the stream orbit in the average $g_z$,
we obtain
\begin{equation}
{{d\langle\omega^2_z\rangle}\over{dt}}\approx {{16\pi^{3/2}}\over{\sqrt{3}}} \ln\Lambda {{G^2 \rho^2(r_c)} \over{v_c^3(r_c)}}  g_z^2\left[r(t)\right] \ r_{sc,1}^3\ m_{22}^{-3}\ ,
\label{thz}
\end{equation}
where $\ln\Lambda$ is the classical Coulomb logarithm, which we estimate in Sect.~3 using a suite of numerical experiments.
As expected for a diffusive process, the vertical angular thickness grows approximately like $\langle\omega_z^2\rangle\propto t$.

Given the stream orbit $r(t)$, a model for the gravitational potential of the Milky Way and a probabilistic description of the properties of FDM clumps, \
the dimensionless quantity $g_z$ can be calculated exactly. We use the working assumptions listed in Sect.~2.1 and
consider the case in which the underlying gravitational potential and time-averaged dark matter density distribution are approximately
scale-free. In this case $g_z$ does not depend on the orbital energy of the stream, but only on its circularity $j$. 
Here $j\equiv J/J_c(E)$, where $J$ is the angular momentum and $J_c(E)$ is the angular momentum of a circular orbit with the same 
energy $E$, corresponding to the radius $r_c$ and to a circular velocity $v_{c}(r_c)$.

\begin{figure}
\centering
\includegraphics[width=.5\textwidth]{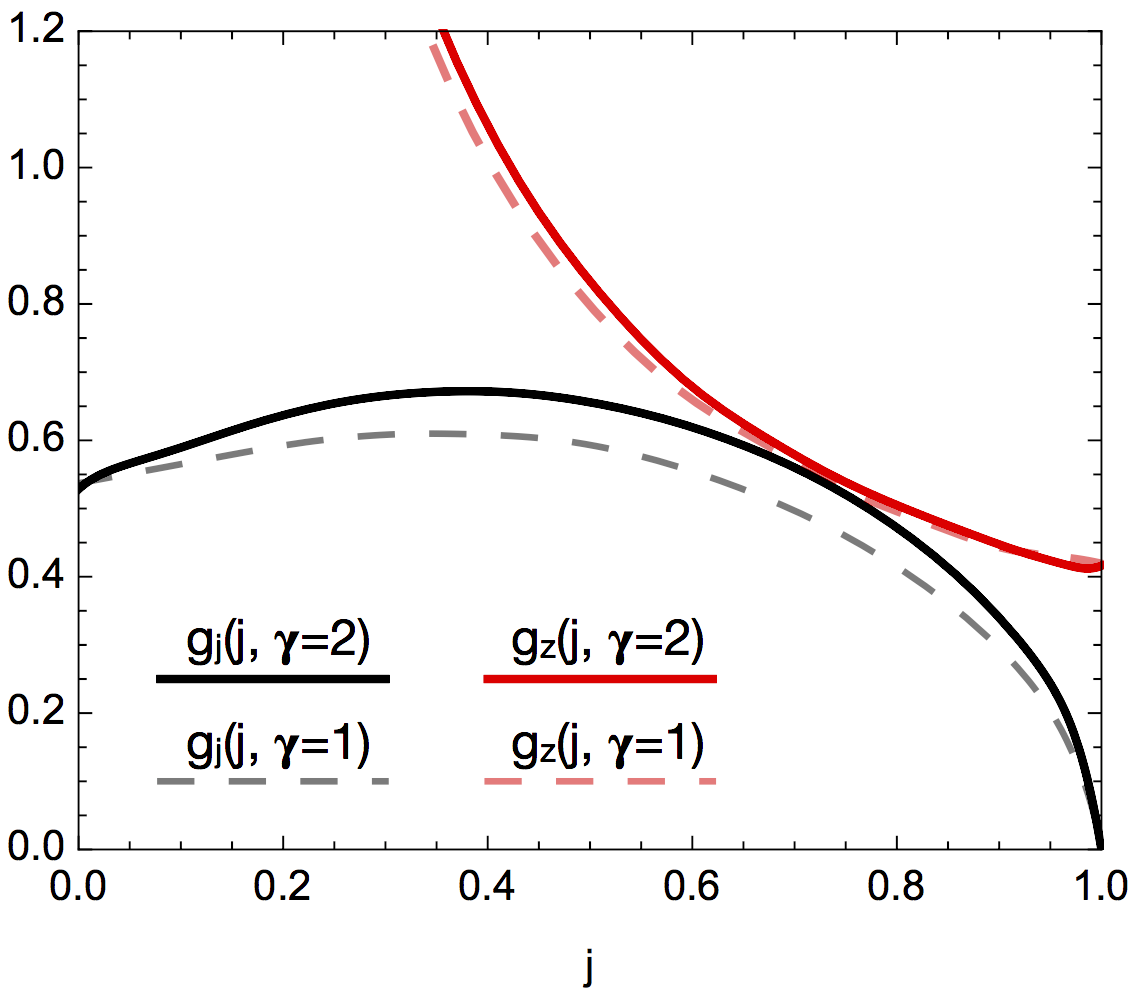}
\caption{\label{gz}: The functions $g_z(j)$ and $g_j(j)$ describing the diffusive thickening of streams in FDM haloes, 
as in Eqs.~(\ref{thz}) and~(\ref{thp}). Full lines refer to the case of streams orbiting an underlying logarithmic spherically symmetric gravitational
potential, corresponding to a power law exponent $\gamma =2$ for the total density profile. Dashed lines refer to the central 
regions of a total density profile with power law index $\gamma =1$, as in the case of an NFW halo.}
\end{figure}

The full red line in Figure~\ref{gz} shows the magnitude of the function $g_z(j)$ for the case of an underlying logarithmic gravitational potential, $\Phi(r)=v_c^2\log (r)$, 
corresponding to a total density distribution $\rho\propto r^{-\gamma}$, with $\gamma=2$. This explicitly uses the properties of orbits in such a potential,
while assuming that the distribution of clump velocities $f(r,\mathbf{v_{lc}})$ is Maxwellian and isotropic, with constant 1-dimensional velocity dispersion 
$\sigma_{cl}=v_{c}/\sqrt2$.
The resulting $g_z$ is a monotonic function of $j$, and increases for streams on eccentric orbits. This should be expected as, all the rest being the same,
encounters close to orbital apocenter are more effective at tilting a star's orbital plane, as can be gathered by Eqn.~(\ref{ztilt}). 
We also consider the case $\gamma=1$, describing the inner regions of an NFW density profile,
and show the corresponding function $g_z$ with a red dashed line in the same Figure. It appears that the average $g_z$ 
is only marginally sensitive to the detailed properties of the underlying potential. \\

The thickening captured by Eq.~(\ref{thz}) is cumulative: regions of the stream that are further away from the progenitor
have been shedded at an earlier time and are therefore predicted to have grown hotter and thicker after interacting with the
quantum clumps for longer periods of time. If we consider the stars at a distance $l$ along the stream from the progenitor GC 
and assume these have been orbiting freely for a time $t=t(l)$ after being tidally stripped, 
the vertical angular thickness at that location in the stream is:
\begin{equation}
\langle\omega_{z}^2\rangle^{1/2}(l)=\left(\omega^2_{z,0}+{{d\langle\omega^2\rangle}\over{dt}} t\right)^{1\over 2}\ . 
\label{lthz}
\end{equation}
Here, $\omega_{z,0}$ is the vertical angular width of an unperturbed stream, which is constant in time if the background potential is spherically
symmetric. 

Since the growth in the length of the stream is secular, $l(t)\propto t$, the effect of FDM turbulence should manifest itself in 
streams with flaring vertical density profiles, i.e. with a vertical thickness that increases while moving away from the 
remnant of the progenitor, with a behavior $\omega_z (l)\propto l^{1/2}$.
In practice, a prediction for the local angular width as function of position along the stream is especially challenging, 
first of all because the location of the remnant itself is unknown for most Milky Way GC streams. Additionally, a detailed estimate of the 
stream age as a function of location requires a precise knowledge of GC mass and internal kinematics, stream orbit and Milky Way potential,
all of which influence the speed at which the stream grows in length \cite{JohnstonS,AmoriscoS,ErkalNS}.
For these reasons, we also estimate a mean global angular thickness for the entire stream, $\langle\omega_z^2\rangle^{1\over2}$, averaged over its full length:
\begin{equation}
\langle\omega_z^2\rangle^{1\over2}(t)=\left(\omega^2_{z,0}+\Gamma_z{{d\langle\omega^2\rangle}\over{dt}} t\right)^{1\over 2}\ , 
\label{mthz}
\end{equation}
The magnitude of the coefficient $\Gamma_z$ depends on the time dependence of the mass loss rate of the parent GC.
Assuming this is approximately constant in time \cite{Renaud, Kupper} gives us  
\begin{equation}
\Gamma_z\equiv \int^1_0 \lambda^2 d\lambda = {1\over{3}} \ .
\end{equation}
\subsubsection{Angular width in the orbital plane}

In the plane of the orbit, the angular width of a stream is due to the spread in the angular momentum of its stars \cite{JohnstonS,AmoriscoS}.
Stars move away from the progenitor GC at a speed that depends on their energy,
so that stars at a specific location along the stream have similar orbital energies \cite{JohnstonS,AmoriscoS}. The angular width of the stream 
in the orbital plane therefore reflects the spread in orbital circularities: the angle between 
two different stars with similar energy but differing by $\delta j$ is
\begin{equation}
\delta\phi\approx{{\partial \phi} \over{\partial j}}\ \delta j\ {t\over T_r} \ ,
\label{secular}
\end{equation}
where $T_r$ is the radial period of the stream and $\phi$ is the corresponding azimuthal angle. Note that 
such angular distance grows secularly, implying that even when unperturbed
the stream phase mixes with time within the orbital plane. 

To add to this, scattering with the quantum clumps causes the spread in circularities of stars in the stream to progressively grow with time. 
Assuming again the number of encounters is sufficiently large, the relevant diffusion is  
\begin{equation}
{{d\langle\delta j^2\rangle}\over{dt}}\approx\left\langle {\iint} f(r,\mathbf{v_{cl}})d^3\mathbf{v_{cl}}  \ 2\pi b db\ n\ v_{rel}\ \delta j^2(\mathbf{\delta v})\right\rangle_{r(t)}
\end{equation}
where $\delta j(\mathbf{\delta v})$ is the change in circularity caused from the velocity change $\mathbf{\delta v}$. 
The change $\delta j(\mathbf{\delta v})$ depends on the clump velocity $\mathbf{v_{cl}}$ and on the specific location 
of the encounter, through the orbit of the stream $r(t)$. As for Eqn.~(\ref{thz}), we use the impulse approximation 
and absorb all dependences on the stream orbit in an average $g_j$, to get
\begin{equation}
{{d\langle\delta j^2\rangle}\over{dt}}\approx {{16\pi^{3/2}}\over{\sqrt{3}}} \ln\Lambda {{G^2 \rho^2(r_c)} \over{v_c^3(r_c)}}  g_j^2\left[r(t)\right] \ r_{sc,1}^3\ m_{22}^{-3}\ . 
\label{thp}
\end{equation}
Figure~\ref{gz} shows the magnitude of the function $g_j(j)$ for the cases of an underlying logarithmic gravitational 
potential and for the case of a cuspy NFW total density profile.

As for the stream's effective vertical velocity dispersion, the stream's spread in circularity also grows as expected for a diffusive 
process: $\langle\delta j^2\rangle\propto t$.
However, the coupling with the intrinsically secular dynamics of the stream captured by Eqn.~(\ref{secular})
causes the stream thickness in the orbital plane to grow in a quicker fashion\footnote{Deviations from spherical symmetry in the background 
potential introduce an analogous coupling in the evolution of the vertical thickness, but we do not consider this effect here.}. 
At a location $l$ along the stream, composed of stars that have been tidally lost at a similar time $t(l)$
in the past, the local angular thickness in the orbital plane is 
\begin{equation}
\langle\omega_{\parallel}^2\rangle^{1/2}(l) \approx \left[\omega^2_{\parallel,0}(t)+\Gamma_l \left({{\partial \phi}\over {\partial j}}\right)^2{{d\langle\delta j^2\rangle}\over{dt}}{{t^3}\over T^2_r}\right]^{1\over2}\ ,
\end{equation}
The coefficient $\Gamma_l$ accounts for the formal convolution of both thickening effects: the spread in circularity grows diffusively thanks to the FDM clumps, 
and this growth is then secularly amplified by the stream's own internal dynamics:
\begin{equation}
\Gamma_l\equiv \int^1_0 (1-\lambda)\lambda^2 d\lambda = {1\over{12}} \ .
\end{equation}
When orbiting in a smooth potential the angular thickness in the orbital plane should flare like $\omega_{\parallel,0}(l)\propto l$,
due to inevitable phase mixing. The quicker flaring, $\omega_{\parallel}(l)\propto l^{3/2}$, testifies 
the presence of a turbulent density field.
Averaging over the full stream of age $t$, we get a global mean angular width of
\begin{equation}
\langle\omega_{\parallel}^2\rangle^{1\over2}(t) = \left[\omega^2_{\parallel,0}(t)+ {\Gamma_\parallel\over12} \left({{\partial \phi}\over {\partial j}}\right)^2{{d\langle\delta j^2\rangle}\over{dt}}{{t^3}\over T^2_r}\right]^{1\over2}\ ,
\label{mthp}
\end{equation}
where, due to the different dependence with time/location, $\Gamma_\parallel \neq \Gamma_z$, and, for a constant mass loss rate, $\Gamma_\parallel=1/4$.

In the following, we make the conservative choice of taking a constant $\omega_{\parallel,0}$, 
as, in practice, its growth rate depends critically on the poorly known mass of the progenitor GC and on the 
detailed properties of the Milky Way potential \cite{AmoriscoS}.
We calibrate both vertical and parallel unperturbed widths using a suite of numerical simulations of perturbed and 
unperturbed streams in Sect.~3, where we also show that a Coulomb logarithm of magnitude $\ln\Lambda\approx9.6$, constant in time,
well describes the evolution of streams following a variety of orbits in our numerical experiments. 

\subsubsection{Magnitude of the heating effect}
We use the model just described to illustrate with an order of magnitude estimate the importance of 
dynamical heating for GC streams in the Milky Way. For a dark matter density of $\rho_\odot=10^7
M_\odot$~kpc$^{-3}$ at the solar radius $R_\odot$, a circular velocity of $v_{c}=220~$km~s$^{-1}$ and the most conservative case of a stream on a
circular orbit, $j=1$, Eqn.~(\ref{thz}) predicts a vertical thickening rate of 
\begin{equation}
\left.{{ d\langle\omega^2_z \rangle } \over {dt}}\right|_\odot \approx\left(0.2~{{\rm deg}}\times  {R_\odot\over r_c}\times m_{22}^{-1.5}\right)^2\bigg/{\rm Gyr} \ ,
\label{mag1}
\end{equation}
corresponding to an increase in the effective vertical velocity dispersion of the stream of
\begin{equation}
\left.v^2_c{{ d\langle\omega^2_z \rangle } \over {dt}}\right|_\odot\approx\left(0.8~{{{\rm km~s}^{-1}} }\times  {R_\odot\over r_c}\times m_{22}^{-1.5}\right)^2\bigg/{\rm Gyr}\ ,
\label{mag2}
\end{equation}
These figures are comparable in magnitude with the observed properties of thin GC streams in the Milky Way 
\cite{OdenkirchenW,OdenkirchenK,Koposov,Grillmair4,Sesar}, showing that cold stellar stream can indeed provide
a useful probe in the interesting range of axion masses. In fact, depending on the viewing angle of the stream, 
the observed thickening may be substantially larger, due to the different scaling of Eqns.~(\ref{mthz}) and~(\ref{mthp}).  
As mentioned in Sect.~2, heating is also predicted to become stronger towards the central regions of the Galaxy, 
where the local dark matter density increases, implying a corresponding increase in the mass of the quantum clumps, Eqn.~(\ref{meff}).
Assuming the density profile of the Milky Way is well described by an NFW profile results in the inverse scaling with 
radius of Eqns.~(\ref{mag1}) and~(\ref{mag2}). We come back to this in Sect.~5.

\begin{figure*}\centering
\includegraphics[width=.88\textwidth]{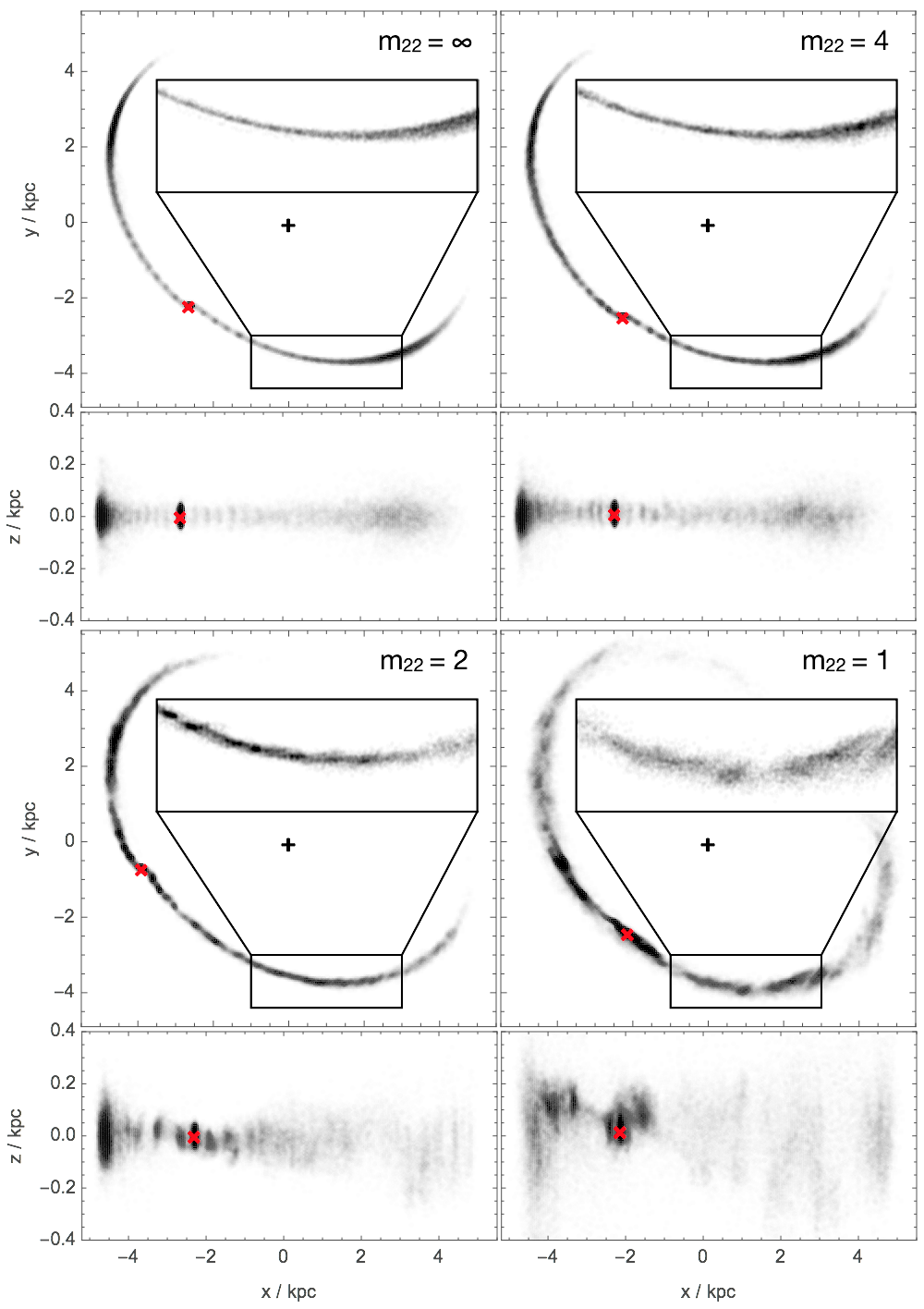}
\caption{\label{puppets} The density distributions of identical GC streams ($r_c=4.4~$kpc, $j=0.9$) evolved for 3 Gyr in 
Milky Way sized FDM haloes with decreasing axion mass. Upper panels show projections in the plane of the orbit, 
lower panels show vertical projections.The red cross indicates the position of the remnant.}
\end{figure*}

\section{Numerical examples}

We use a suite of N-body simulations to exemplify the heating of thin streams in FDM haloes
and to test the analytic description developed in Sect.~2. Our simulations are based on the publicly available 
N-body code {\it Gadget 2} \cite{SpringelG}, which we however modify to introduce the phenomenological 
quasiparticle model described in Sect.~2.1. 

\subsection{Effective FDM clumps in \it{Gadget 2}}

In each of our effective N-body simulations, we follow the evolution of a GC and its stream for 10 Gyr.
We inizialize the GC so that its orbital plane is $z=0$, and its angular momentum is $\mathbf{J_{GC}}=J_{GC}\mathbf{\hat{z}}$.
Since we do not wish to use these simulations to directly reproduce the properties of observed streams, 
we can adopt a simplified mass model for the Galaxy. We assume that the total background potential 
follows a spherically symmetric NFW density distribution, with dimensional scalings appropriate 
for a Milky Way sized dark matter halo (characteristic radius $r_0=20.6~$kpc, characteristic 
density $\rho_0=6.1\times10^6M_\odot$~kpc$^{-3}$).
This is the same as the time-averaged density distribution associated to the quantum clumps themselves.  
 
In order to keep the number of `live' quasiparticles and the computational effort reasonable, 
we represent the gravitational field of the FDM halo using the combination of a static, smooth background 
gravitational potential and live quantum clumps. The static gravitational potential is the one appropriate for the mass model
of the Galaxy above. Then, we add a population of live quasiparticles, but only do so in the vicinity of the GC stream. 
In practice, at any single time, our simulations include live quasiparticles in the region $|z|<z_{max}$ (and $r<r_{max}$),
where we take care to fix $z_{max}$ (as well as $r_{max}$) such that this remains several times wider than the stream itself, 
during its entire evolution. This keeps the number of quasiparticles manageable and allows us to neglect the `additional' mass 
introduced by the live clumps, which is always negligible with respect to the one 
imposed by the static potential, which drives the undisturbed dynamics of the stream.

According to the prescription of Eqn.~(\ref{meff}), clumps have a uniform number density, $n=1/V_{cl}$,
which we use to generate the initial conditions of our live quasiparticles, randomly filling the volume within the boundary $|z|<z_{max}$. 
Eqn.~(\ref{meff}) also prescribes that the mass of individual quasiparticles varies with radial location. For simplicity,
we force our effective clumps to move on circular orbits, which keeps their individual mass constant in time.
When the dynamics of individual quantum clumps is better resolved in self-consistent FDM simulations, this may be changed. 
For now, we assume quantum clumps move on randomly oriented orbital planes, with constant circular velocities, 
sampled from a Maxwellian distribution with dispersion that depends on radius, tracking the velocity dispersion of an isotropic dark matter halo 
with the same density profile. 
When individual quasiparticles cross the boundary $|z|=z_{max}$ and move out of the turbulent region, we individually replace them.
We do so by introducing a new live clump which enters the turbulent region: we sample a new orbital radius together with new direction
and magnitude for its angular momentum. This ensures that number density of live quasiparticles remains constant 
as well as approximately uniform during the entire simulation.

Within this strategy it is easy to track the location of each live quasiparticle in the simulation, analytically. 
We modify {\it Gadget 2} to do so and to calculate the perturbing force that each clump exerts on the stars
that compose the GC and stream. We recall that we assume that effective clumps are represented by Plummer spheres, 
with a half-mass radius $\epsilon=r_{sc}$.

\begin{figure*}
\centering
\includegraphics[width=.88\textwidth]{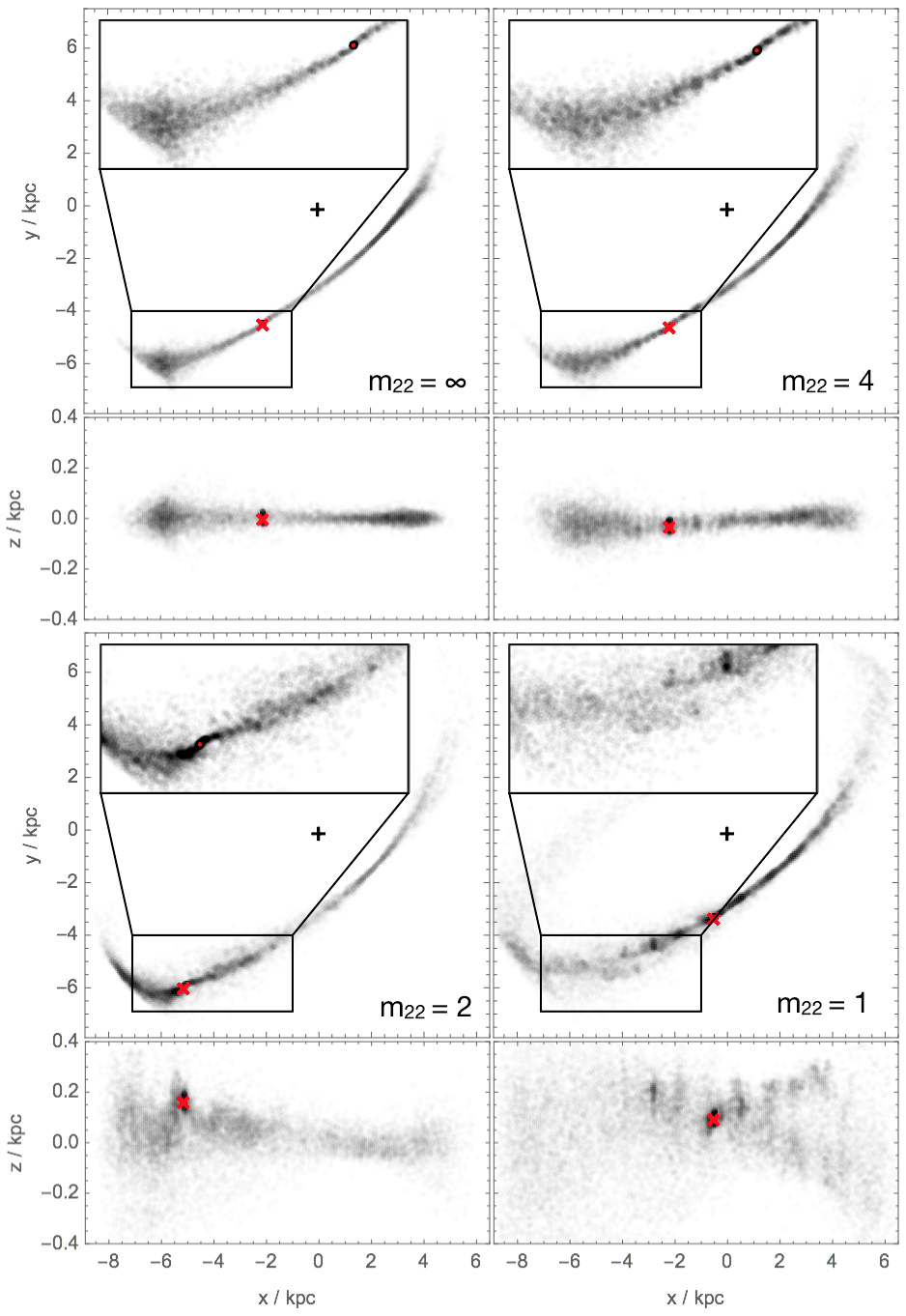}
\caption{\label{puppetsecc} The density distributions of identical GC streams ($r_c=6.3~$kpc, $j=0.6$) evolved for 4.9 Gyr in 
Milky Way sized FDM haloes with decreasing axion mass. Upper panels show projections in the plane of the orbit, 
lower panels show vertical projections.The red cross indicates the position of the remnant.}
\end{figure*}

\subsection{The suite of runs}

In each run, the progenitor GC is represented by a live isotropic Plummer sphere, sampled with $10^5$
equal mass particles. We explore a sample of different properties for the progenitor GC: we consider
progenitors with masses in the range $4\leq\log M_{GC}/M_\odot\leq 4.8$, and 
initial half-mass radii in the range $6\leq r_{eff}/{\rm pc}\leq 16$. We also explore different orbits.
We consider the case of an almost circular orbit, circularity $j=0.9$, and the case of a significantly 
eccentric orbit $j=0.6$. For the former, we consider two different orbital energies,
$r_c=7.8~$kpc, corresponding to a stream that orbits close to the solar radius, and $r_c=4.4$~kpc, corresponding to 
a stream that inhabits the central regions of the Galaxy. The orbital energy of the eccentric case 
is $r_c=6.3$~kpc. For each of these three orbits we simulate multiple instances of the cases 
$m_{22}=1$, $m_{22}=2$, $m_{22}=4$.

Figure~\ref{puppets} shows the density distributions of otherwise identical streams evolved in FDM haloes with different $m_a$, at $t=3$ Gyr. 
The displayed snapshots refer to the case of GCs on the orbit $(j, r_c)=(0.9, 4.4~{\rm kpc})$, with identical progenitors with $\log M_{GC}/M_\odot=4$ and 
$r_{eff}=6~{\rm pc}$. The case $m_{22}=\infty$ shows the case of an unperturbed stream, orbiting only under the 
influence of the smooth and static background potential, with no live clumps. For decreasing values of 
$m_{22}$ streams become gradually kinematically hotter and thicker. 
Figure~\ref{puppetsecc} shows similar snapshots for streams generated by progenitor GCs with the same properties, 
but on a different orbit, with energy $r_c=6.3 kpc$ and significantly more eccentric $j=0.6$. The snapshots
show the stream after $t=4.9$~Gyr of evolution.

Figs.~\ref{puppets} and~\ref{puppetsecc} also
show that streams become increasingly disturbed and structured with decreasing axion mass. 
This applies to both their density distributions and 
kinematics. This behavior is due to the quickly increasing mass of 
individual quantum clumps, Eqn.~(\ref{meffr}), as well as to number of individual encounters experienced by the stream,
which decreases at the same rate. When axion masses are $m_{22}\lesssim1$ individual encounters
may be effective enough to cause the formation of gaps and kinks in the stream, as
for the case of dark matter subhaloes and giant molecular clouds. Differently from what happens for 
higher axion masses, these encounters shape the structure of the stream
with individual features. These are more persistent when $m_{22}\lesssim1$ as 
longer times are required for subsequent encounters to erase them, due to the reduced rate of flybys.

These `large-scale' perturbations are not explicitly predicted by our diffusive model, which is based on the assumption 
of a large number of uncorrelated encounters. Rather than causing displacements in the mean location of the 
stream, uncorrelated encounters simply increase its width in phase space. 
However, as we show in the following Section, we still find that our analytic model provides a good description 
of the evolution of the average width of the stream, at least within the range of axion masses we explore, $m_{22}\geq1$.

\begin{figure*}
\includegraphics[width=\textwidth]{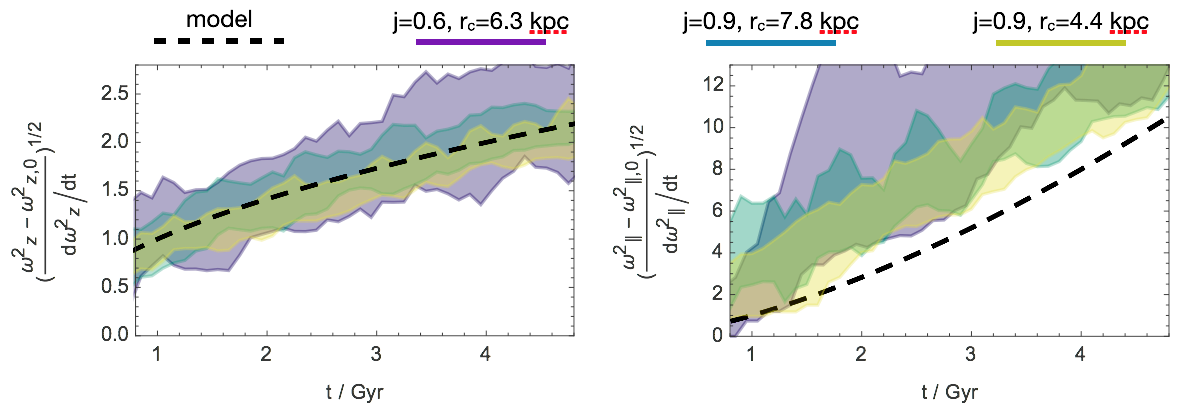}
\caption{\label{test} A comparison between our analytic model for the stream thickening (black dashed lines) and 
results from our suite of numerical experiments. Shaded areas (10-to-90\% quantiles) pertain numerical results for 
sets of streams with the same orbital properties (as in the legend). Different axion masses and progenitor properties 
are explored for each orbit.}
\end{figure*}

\subsection{Tests of the analytic model}

Figure~\ref{test} compares the global mean angular width of streams in our numerical suite with the predictions of our analytic model.
The left panel pertains to the vertical angular width, and displays results in terms of the quantity $\left({{\omega^2_z-\omega^2_{z,0}}\over{d \omega_z^2/dt }}\right)^{1\over2}$.
Streams with different progenitors and orbits have different constant undisturbed widths $\omega^2_{z,0}$. 
Different axion masses as well as stream orbits imply different growth rates $d \omega_z^2/dt$. However, according to Eqn.~(\ref{mthz}),
our model predicts that this combination should scale simply as
\begin{equation}
\left({{\omega^2_z-\omega^2_{z,0}}\over{d \omega_z^2/dt }}\right)^{1\over2} = t^{1\over2} \ .
\end{equation}
The black dashed curve in left panel of Fig.~5 displays such scaling. Shaded areas are simulation results. They group all streams 
in our suite with the same progenitor's orbit, and extend between the 10\% and 90\% quantiles over the different set of realizations. 
We measure the mean vertical angular width using all particles that are no longer bound to the progenitor GC,
and calculating the standard deviation of the distribution of individual inclinations $\omega_z=\arctan\left( z/\sqrt{x^2+y^2}\right)$. 

We use a simple model for the unperturbed vertical width 
\begin{equation}
\omega_{z,0} = 
k_z \left[{{M_{GC}}\over{M_{tot}(r_c)}}\right]^{1\over 3} \left(r_c\over r_p\right)^{1\over 2} \ ,
\label{basew}
\end{equation}
where $r_p$ is the pericentric radius of the stream orbit. We calibrate the dimensionless 
coefficient $k_z$ on our numerical results, finding $k_z=0.45$. This simple model 
well reproduces the undisturbed, constant vertical angular width of streams with different properties.
As mentioned earlier, we also use the comparison in Fig.~\ref{test} to calibrate the magnitude of the 
Coulomb logarithm, which we fix at $\ln\Lambda=9.6$. As shown by the same panel, we find that a constant Coulomb logarithm well describes 
the evolution of our disturbed streams\footnote{We would expect the Coulomb logarithm to grow in magnitude as streams become
thicker with time, but our simulations do not suggest that this adjustment is needed in the regime we are exploring here. }. 
Shaded areas closely follow the predicted dependence with time, 
highlighting how a diffusive description of the heating process is appropriate for the evolution of the vertical thickness.

The right panel of Figure~5 pertains to the thickness of the stream in the orbital plane. 
We measure the mean angular width in the plane of the stream as the standard deviation of the distribution of angular distances from 
the centerline of the stream (defined by a spline interpolation), over all particles that are no longer bound to the progenitor GC. 
Similarly to the left panel, numerical results are displayed in terms of the combination $\left({{\omega^2_\parallel-\omega^2_{\parallel,0}}\over{d \omega_\parallel^2/dt }}\right)^{1\over2}$ and we use the same Coulomb logarithm.
Furthermore, we take $\omega_{\parallel,0}=\omega_{z,0}$, as defined in Eqn.~(\ref{basew}), and ignore that the unperturbed angular width 
in the orbital plane evolves linearly with time. This causes a departure between the predicted width
\begin{equation}
\left({{\omega^2_\parallel-\omega^2_{\parallel,0}}\over{d \omega_\parallel^2/dt }}\right)^{1\over2}=t^{3\over 2}\ ,
\end{equation}
and what observed in the numerical experiments: shaded areas lie systematically above our model prediction, 
especially for the case of eccentric streams, which have a comparatively faster unperturbed growth rate.
We are however not worried by this as this simplification makes our analytic model inherently conservative. 

\section{First constraints on the axion mass}

We use our analytical model to provide a first early constraint on the axion mass. 
In the presence of dark matter subhaloes or of additional baryonic perturbations 
streams will thicken beyond our model predictions. Therefore, our model quantifies the minimum 
width of streams in FDM haloes, which can be used to establish a lower limit for the axion mass $m_{22}$.

We use pre-Gaia literature data for six Milky Way streams: Palomar~5 \cite{OdenkirchenK}, 
Ophiucus \cite{Sesar}, GD1 \cite{Koposov}, Acheron, Cocytos and Lethe \cite{Grillmair4}, and we collect adopted values in Table~1.
 \begin{table}
\caption{\label{table}  Properties adopted for the cosidered Milky Way streams.}
\centering
\begin{tabular}{cccccc}
\hline
\textrm{Stream}&
\textrm{pericenter }&
\textrm{apocenter }&
\textrm{prog. mass}&
\textrm{sky-projected}&
\textrm{Refs.}\\

 & [kpc] & [kpc] & $\log M_{GC}/M_\odot$ &\textrm{width} & \\
\hline

Palomar 5 & $7\pm1$ & $19\pm0.5$ & 4 & FWHM$=120$ pc  & \cite{OdenkirchenW,Kupper,ErkalNS}\\
GD1 & $13\pm1$ & $22\pm2$ & 4 & $\sigma=12'$  & \cite{GrillmairD, Koposov, ErkalNS}\\
Ophiucus & $3.6\pm0.1$ & $16.8\pm0.5$ & 3.4 & $\sigma=4'$  & \cite{Bernard, Sesar}\\
Acheron& $3.5\pm0.8$ & $9.2\pm3.3$ & 4 & FWHM$=60$ pc  & \cite{Grillmair4}\\
Cocytos& $4.9\pm0.2$ & $12.5\pm0.2$ & 4 & FWHM$=140$ pc  & \cite{Grillmair4}\\
Lethe& $7.7\pm0.4$ & $17.3\pm0.5$ & 4 & FWHM$=95$ pc  & \cite{Grillmair4}\\
\hline
\end{tabular}

\end{table}
For each stream, we compare the observed angular width with what predicted by our model as a function of the axion mass,
and we interpret this comparison as a bound on $m_{22}$ using the following model likelihood:
\begin{equation}
\mathcal{L}(m_{22})=\left\{
\begin{array}{lr}
\exp \left[-{1\over2}  \left(\left({\omega_{obs}-\omega_{mod}}\right)/{\Delta\omega}\right)^2\right] & \hspace{1cm} {\rm if\ } \omega_{obs}<\omega_{mod}\\
1& {\rm otherwise,}
\end{array}\right. \ .
\label{lik}
\end{equation}

The model angular thickness $\omega_{mod}$ depends on a number of nuisance astrophysical parameters.
First, we require a model for the Galactic potential. 
As suggested by the approximately constant rotation curve of the Galaxy in the region of interest \cite{BlandHawthorn,Sysoliatina},
we adopt a logarithmic total gravitational potential, and take that the time-averaged dark matter density distribution follows a NFW density profile.
We assume uncorrelated normally-distributed probability distributions for the following nuisance parameters and
marginalise over them: 
{\it (i)} total circular velocity and solar radius, respectively $v_{c}=220\pm 10~$km~s$^{-1}$ and $R_\odot=8.25\pm0.2~$kpc \cite{BlandHawthorn,Bovy};  
{\it (ii)} local dark matter density and concentration of the dark halo, respectively $\rho_\odot=10^{7\pm0.11} M_\odot$~kpc$^{-3}$ and $c=10\pm1$
 \cite{Read,Maccio,Ludlow}.

Second, we require orbital properties for the different streams as well as structural properties for their progenitors.
Pericentric and apocentric radii of the different streams are relatively well measured: we use these together with the mass model above
to produce probability distributions for the circularity $j$ and circular radius $r_c$ for each stream.
Obtaining estimates of the progenitor GC mass and age is more challenging. We remain conservative on both these quantities: 
(i) we adopt the progenitor masses listed in Table~1 and (ii) we estimate stream ages based on the lengths of the observed (section of) each stream.
We get respectively 3, 5, 1.5, 5, 6, 1.5 Gyr for Palomar~5,  GD1, 
Acheron, Cocytos and Lethe and Ophiucus, in line with other estimates in the literature \cite{Kupper,ErkalNS}.
Finally, in addition to marginalizing over the astrophysical parameters listed so far, we also account for any
additional uncertainty -- in either the observed width and/or our model prediction -- using a standard deviation of 
$\Delta\omega=0.15\times\omega_{obs}$ when computing the likelihood~(\ref{lik}).

\begin{figure*}
\includegraphics[width=\textwidth]{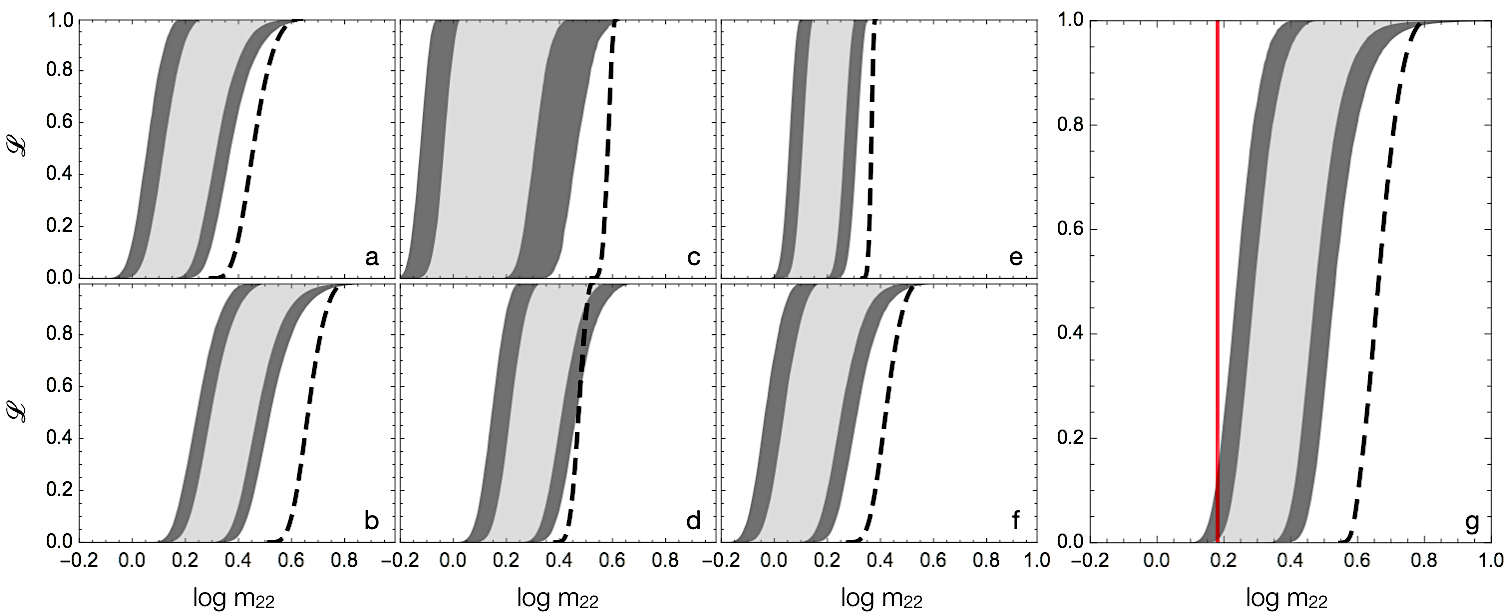}
\caption{\label{limit} Constraints on the axion mass $m_{22}$ from the width of individual Milky Way streams (respectively
Palomar~5, Ophiucus, Acheron, Cocytos, Lethe and GD1 in panels {\it a} to {\it f}). Panel ${\it g}$ displays the joint constraint, 
corresponding to a lower limit of $\log m_{22}>0.18$, as shown by the red vertical line.}
\end{figure*}

Use of an individual set of values for the astrophysical parameters and Eqn.~(\ref{lik}) provides a profile for the 
likelihood $\mathcal{L}(m_{22})$ for each of our six streams, and therefore an exclusion region for
the axion mass. By sampling the probability distributions of all nuisance parameters, we construct 
the shaded areas in the panels {\it a} to {\it f} in Figure~\ref{limit}.
Each panel displays the 1-$\sigma$ and 2-$\sigma$ regions for the profiles of the likelihood $\mathcal{L}$, as a function of the axion mass $m_{22}$.
From {\it a} to {\it f}, the different panels show our inference for Palomar~5, Ophiucus, Acheron, Lethe, Cocytos and GD1, respectively. 
Panel {\it g} shows the combined inference, assuming uncorrelated measurements. As we cannot attribute the thickness of the observed streams to 
FDM turbulence alone, the upper bounds of the shaded areas in Fig.~\ref{limit} cannot be interpreted as upper bounds for the axion mass. 
However, as any additional form of heating would be simply added to the one caused by FDM, Fig.~\ref{limit}
provides valid lower bounds. 

To derive nominal lower bounds for the axion mass we take the full sample of joint likelihood profiles $\mathcal{L}(m_{22})$
in panel {\it g} and, at any fixed $m_{22}$ consider the top 95\% quantile. This defines an exclusion region that is 
valid for 95\% of all sets of nuisance parameters. For this choice, a 2-$\sigma$ lower limit corresponds to 
$\log m_{22}>0.18$, as shown in panel {\it g} by a vertical red line. Use of the likelihood profile 
$\mathcal{L}(m_{22})$ corresponding to the top 84\% quantile over all sets of nuisance parameters together 
with a 1-$\sigma$ exclusion would give us a tighter limit of $\log m_{22}>0.3$.

Reduced uncertainties on the properties of the considered streams as well as on the other astrophysical parameters will
allow for tighter constraints in the future. For instance, the total length of several of the used streams is currently unknown, 
as they appear truncated by the sky coverage of the survey in which they have been first discovered \cite{GrillmairD,Grillmair4}. 
The dashed lines in Figure~2 show profiles for the likelihood~(9) that result from less conservative assumptions on the ages of our streams  
(respectively 4, 9, 4, 8, 9, 4 Gyr for Palomar~5,  GD1, Acheron, Cocytos, Lethe and Ophiucus). These likelihood profiles
adopt the fiducial values for the reported astrophysical parameters and an uncertainty of 3\% for the 
observed average thickness of each stream, $\Delta\omega$. As shown in panel {\it g}, this would correspond to a joint 2-$\sigma$ 
lower limit of $\log m_{22} > 0.6$. Parallel use of kinematic data would also strengthen results.

\section{Discussion and Conclusions: an independent dynamical probe of FDM} 

We have shown that the turbulent behavior of FDM haloes produces observable 
dynamical heating of the kinematically cold stellar streams of GCs. 
Using a phenomenological description of the fluctuating dark matter density field, we have 
shown that this effect can be important for thin stellar streams in the Milky Way and that it is potentially observable
for the range of axion masses that is interesting for cosmology. This shows the 
promise for obtaining independent local dynamical constraints to FDM.

Currently available constraints to the axion mass are all similarly derived from modelling the power spectrum of the Lyman-$\alpha$ forest \cite{Irsic,Armengaud,KobayashiL}. These studies consistently appear to disfavour the range of axion masses that would result in significant 
astrophysical implications. Taken at face value, these studies rule out FDM models with $m_{22}\lesssim20$, 
entirely excluding the window of boson masses required to alleviate the small-scale tensions with the CMD model.
Since these results all rely on the same physics and on similar modelling, it is important to identify different, 
independent probes of the axion mass. It has been claimed, in fact, that a number of astrophysical effects may 
cause the constraints above to be substantially relaxed, perhaps by an order of magnitude \cite{Hui}. 
Based on entirely different physics, the perturbation and heating of GC streams in the Milky Way can provide us 
with such a fully independent test of the FDM model.

For this purpose, we have constructed an analytic model for the effects of quantum turbulence on the stream thickness.
This is based on a quasiparticle effective model of the turbulent density field of FDM haloes.
We find that streams experience a process of diffusive heating, 
causing their internal velocity dispersion and vertical angular width to grow as $t^{1/ 2}$. Because 
of the secular dynamics of the stream, the evolution of the angular width in the orbital plane is faster. 
We have tested these scalings using a suite of tailored numerical simulations and found that our
analytic model well reproduces the process of diffusive dynamical heating caused by the random 
encounters with the FDM clumps.

\begin{figure}
\centering
\includegraphics[width=.5\columnwidth]{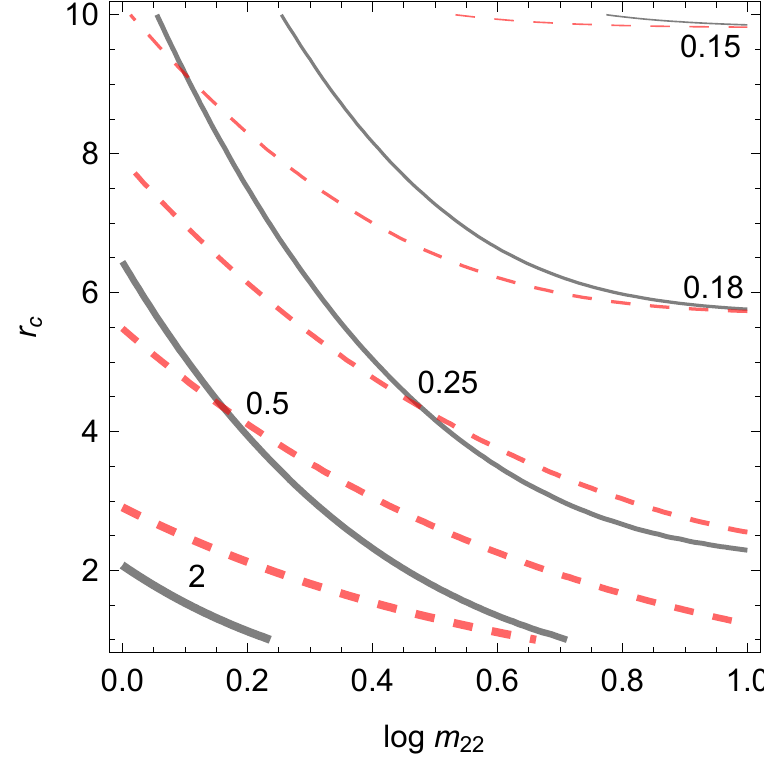}
\caption{\label{in}  The minimum angular width in degrees for young streams 
(age $t=2$~Gyr) with progenitors of mass $\log \left(M_{GC}/M_\odot\right)=4$ and orbital circularity $j=0.85$.
The widths in the direction perpendicular 
and parallel to the orbital plane are displayed by black full lines and red dashed lines.
}
\end{figure}

Using conservative assumptions on the properties of six Milky Way streams and pre-Gaia literature data, 
we have derived a first lower limit of $m_{a}>1.5\times10^{-22}$~eV. Improved data for the same streams and tighter constraints
on the gravitational potential of the Galaxy will likely allow for a correspondingly tighter limit. 
Furthermore, the detection of previously unknown GC streams in the central regions of the Galaxy 
may also allow new stringent limits. Figure~\ref{in} shows the minimum angular width (in degrees)
predicted by Eqs.~(\ref{mthz}) and~(\ref{mthp}) for young streams (age $t=2$~Gyr) with progenitors 
of mass $\log \left(M_{GC}/M_\odot\right)=4$ and almost circular orbits ($j=0.85$),
for a range of orbital energies and axion masses. The detection of any single stream
that is thinner than predicted here would essentially rule out a significant
range of the parameter space.

In this paper, we have focussed on the readily observable mean angular width and velocity dispersion of streams.
However, future observations of the detailed properties of the stream's density 
and kinematic fields will provide additional constraining power, as the heating process by 
the quantum fluctuations is not fully diffusive, as shown by Figures~\ref{puppets} and~\ref{puppetsecc}. 
In fact, the perturbations caused by FDM turbulence and by low-mass CDM subhaloes may be 
degenerate in some regimes, especially within analyses that concentrate on a single stream. 
Analyses based on a set of streams with different orbital energies will be able to disentangle the two effects,
based on the markedly different dependence on galactocentric distance of the heating caused by FDM turbulence. 

Though guided by the results of current self-consistent FDM simulations, the current model remains 
however quite simplified. Future self-consistent simulations of FDM haloes will be able to better resolve 
the dynamics of individual oscillations in the dark matter density field. 
The analytic framework we have presented can easily be improved to incorporate a more faithful description 
of the properties of real quantum clumps. For instance, a better statistical description of the spectrum 
of effective masses and of their motion would be beneficial, together with a characterization of any secondary 
dependences between the two quantities. We anticipate that the analytic model presented here can be used in combination 
with future numerical results from self-consistent simulations to more accurately predict the effect of dynamical heating.
Thanks to the Gaia mission and upcoming Galactic surveys \cite{Gaia,DESI,LSST,4MOST,WEAVE}, dynamical 
heating of stellar streams in the inner Milky Way provides a promising novel method for setting 
independent constraints on FDM models, with the potential for improved and competitive limits.


\acknowledgments
This work was supported in part by the Black Hole Initiative, which is funded by a JTF grant.

\bibliographystyle{unsrt}
\bibliography{FDMbib}

\end{document}